\documentclass[reprint,amsmath,amssymb,aps]{revtex4-1}
\usepackage{graphicx}
\usepackage{dcolumn}
\usepackage{bm}
\usepackage{multirow}         
\usepackage{graphicx}         
\usepackage{graphics}
\usepackage{hyperref}         

\def\bea{\begin{eqnarray}}
\def\eea{\end{eqnarray}}
\def\be{\begin{equation}}
\def\ee{\end{equation}}
\def\fr{\frac}

\def\la{\label}

\def\le{\left}
\def\ri{\right}
\def\Lc{\Lambda_c}

\def\ac{a_c}
\def\LCDM{$\Lambda$CDM}
\def\eV{\textrm{eV}}

\begin{document}

\title{Testing the nature of Dark Energy with Precision  Cosmological constraints}

\author{Axel de la Macorra}
  \email{macorra@fisica.unam.mx, a.macorra@gmail.com}
\author{Erick Almaraz}
  \email{ealmaraz@estudiantes.fisica.unam.mx}
\affiliation{%
Instituto de F\'isica, Universidad Nacional Aut\'onoma de M\'exico\\
Ciudad de M\'exico, 04510 M\'exico 
}

\date{\today}

\begin{abstract}

We  present a Dark Energy  (DE) model with a sound derivation as a natural extension of the Standard Model  of particle physics with no free parameters and an excellent  fit with current cosmological data  improving  by $21\%$ the   \LCDM\, fit of the  Baryon Acoustic Oscillations (BAO) measurements, specially  designed to determine  the dynamics of DE.
DE corresponds to  the lightest bound state scalar particle $\phi$ with  a potential $V=\Lambda_c^{4+2/3}\phi^{-2/3}$  dynamically  formed  at the condensation energy scale $\Lc$  and  scale factor $a_c$.   The value of $\Lc$, the exponent $n=2/3$, and the initial conditions of $\phi$   are  all derived quantities.
We obtain an exact constraint $a_c\Lc/\textrm{eV}=1.0939\times 10^{-4}$ and a theoretical prediction  $\Lambda_c=34 \,  ^{+16}_{-11}\,\textrm{eV}$, consistent with the best  fit  $ \Lambda_c=44.08\pm 0.27\,\textrm{eV}$.  We test  our model constraint  on $a_c\Lc$  by  allowing  $a_c$ and $\Lc$ to vary independently and  remarkably our prediction  has a relative difference of only 0.2\% with the best fit value. Unlike a cosmological constant $\Lambda$, our DE model predicts the amount of DE and leaves detectable cosmological imprints  at different times and scales at  a background and perturbation level.

\end{abstract}


\maketitle

\section{Introduction}

The mysterious accelerating expansion of the Universe has been well established in the last decade by a large  number of  independent observational experiments  to  unravel the origin of Dark Energy. Among these observations we have the Cosmic Microwave Background Radiation (CMB) \cite{Planck2015}, BAO  and Large Scale Structure (LSS) surveys \cite{GilMarin,Ross,Beutler},  Type Ia Supernovae (SNIa) \cite{Betoule}, and local $H_o$ measurements \cite{RiessH0}. Ambitious projects such as DESI \footnote{\url{http://desi.lbl.gov/}}, LSST \footnote{\url{https://www.lsst.org/}} and Euclid \footnote{\url{http://sci.esa.int/euclid/}}  are scheduled to start operating in the near future. The unprecedented  amount of precise cosmological  data gathered in the last decade allows us to set tight constraints and  discriminate DE models. These recent precision cosmological data, in particular the BAO measurements,   show that our DE model  is dynamically  favoured over   \LCDM\, even tough it has one less free parameter.
The energy density of the Universe at present day is made of  69\% DE, 26\% Dark Matter (DM)  while only 5\% corresponds to the Standard Model (SM) particles consisting principally of photons, neutrinos and ordinary matter. Within the context of general relativity  the  standard model in cosmology ($\Lambda$CDM)  assumes a cosmological  constant $\Lambda$  as DE, constant in  space and time,  and has an excellent agreement with the observations \cite{Planck2015}.  However, there is  no understanding  of  the  origin nor magnitude  of $\Lambda$  and hence of why and when the Universe accelerates \cite{Martin}. This leads to two interesting theoretical (philosophical) problems in $\Lambda$CDM commonly referred to as the ``naturalness''  and  the ``coincidence''  problems. The ``naturalness problem'' requires to fine tune the value of  the energy density $\rho_\Lambda$ to an incredible  one part in  $10^{120}$  at an initial  epoch,  usually taken as  the Planck $M_{pl} =1.9\times 10^{19}\,GeV$ or  the unification $\Lambda_{\textrm{gut}}\simeq 10^{16}\,GeV$ scales (see Fig.(\ref{fig:paper_figure1}\textbf{c})), while the ``coincidence problem''  inquires why  the amount of  $\rho_\Lambda$ is of the same order of magnitude as matter $\rho_m$ precisely at present time. Here we show that our DE   model solves  both problems naturally, since it predicts the values of DE  at the  $\Lambda_{\textrm{gut}}$  scale and  at present time avoiding any fine tuning.

Alternative to $\Lambda$,  scalar fields $\phi$ have been proposed as possible sources to describe DE and a wide
range of models  have been studied in recent years  \cite{PlanckDE,Tsujikawa,Bamba,Copeland,Chen,SmerBarreto}. In particular,  inverse power law (IPL) potentials  $V(\phi)= M^{4+n} \phi^{-n}$ proposed by \cite{Peebles,Ratra,Wetterich}  have been widely investigated \cite{Wang,AxelOtto2002,Ferreira} giving  an  equivalent fit as  \LCDM\, \cite{Alimi}.  The  evolution of the energy density $\rho_\phi=\dot\phi^2/2+V(\phi)$ depends on the parameters $n$, $M$ and the initial conditions of $\phi$. These quantities are  free parameters to be adjusted by the cosmological observations or the choice of  model.  For example, for an IPL potential with $n=1/2$, the evolution of the equation of state (EoS) $w=p/\rho$ close to present time can be a decreasing function  from $w=-0.8$ to $w_o=-0.87$  \cite{Alimi}   assuming  $V$ to be in the tracking regime \cite{Steinhardt}. However, for different initial conditions we can have a growing  EoS from $w\simeq -1$ to $w_o\simeq -0.85$ at present time. Clearly the choice of initial conditions of $\phi$ is important and the current  precision cosmological data, in particular the BAO measurements,  allow us to constrain  the dynamics of DE.

Here we present a Dark Energy model that is a natural extension of the SM (perhaps the  most accurate theory in physics  \cite{PDG2016}) where the DE corresponds to the lightest meson scalar particle $\phi$, a ``dark pion", dynamically formed at late times given by the  scale factor $\ac$ as a result  of the non-perturbative dynamics of a hidden Dark Gauge Group (DG) \cite{ADS,Binetruy,Axel2003}. The scalar field $\phi$  is not a fundamental particle but a composite particle and since its mass arises from the binding energy of the  fundamental interaction of the DG, we refer to it as Bound Dark Energy (BDE).
We obtain a scalar potential $V=\Lambda_c^{4+n}\phi^{-n}$ with $n=2/3$, where  the  value of $n$, $\Lc$, the initial conditions of $\phi$  and the onset of BDE at $\ac$  are  all derived quantities.  Remarkably,  our model has no free parameters and  fits better  the cosmological data than $\Lambda$CDM.
Our DE model  constraints  the two parameters $a_c\Lc/\textrm{eV} = 1.0939\times 10^{-4}$  with best fit values  $\Lc=44.02\,\textrm{eV}$ and $a_c = 2.48 \times10^{-6}$.  We  test our model prediction on $a_c \Lc$  by  allowing $a_c$ and $\Lc$ to vary freely  and independently and we  find remarkable  that the relative difference  between the theoretical prediction with the  best-fit value  is only 0.2\%.

Contrary to the standard  \LCDM\,,  where the cosmological constant  has  an important effect only close to present time  but  is  negligible  at early times ($\Omega_{\Lambda}[z>5]<1\%$), our DE model  has a  rich  structure and  contributes to the evolution of the universe   at very different times and scales, leaving cosmological imprints allowing us to probe its validity. At high energies the DG particles are massless and amount to  43\%  of the energy density of the SM at  the  unification scale $\Lambda_{\textrm{gut}}$. Once the BDE is formed at  $a_c$, the BDE density dilutes rapidly ($\rho_{\mathrm{BDE}}\sim a^{-6}$) impacting  the evolution of matter perturbations for modes $k$ entering around  $\ac$  ($k_c=0.925 \textrm{Mpc}^{-1}$) and enhancing them up to   20\% compared to \LCDM. For $a>\ac$, BDE becomes negligible for a long period of time until recently, when it starts  growing  to finally dominate and accelerate the universe close to present time. The evolution of $\rho_\textrm{BDE}$ at late times has a growing EoS  from $w\simeq -1$ to $w_o=-0.93$. Our BDE model modifies the cosmological distances and  structure growth at late times in a similar but distinguishable form than  \LCDM.

\section{Bound Dark Energy}

The dark energy model presented here introduces a supersymmetric Dark Gauge Group (DG)  $SU(N_c)$ with $N_c=3$ colors and $N_f=6$ elementary massless particles in the fundamental representation \cite{Axel2003,Axel2005}.   
The values of $N_c$ and $N_f$ have the same fundamental status as the gauge groups and number of families of the SM ($SU_{QCD}(N_c=3)\times SU(N_c=2)_L\times U_Y(N_c=1)$ and 3 families) describing the strong (QCD), weak and electromagnetic interactions and they are input parameters not derived from a more fundamental theory.  
At high energies  the  DG particles are weakly coupled  and they contribute   to the total content of radiation of the Universe.
However, at lower energies the strength of the DG interaction increases and the gauge coupling becomes strong at the condensation energy scale $\Lc$ and scale factor  $a_c$. At this scale the fundamental fields of the DG form gauge invariant composite states, dark mesons and dark baryons, which  acquire a non-perturbative mass proportional to $\Lc$. This is similar to the strong QCD force, where the masses of the protons and pions are of the order of the  QCD scale $\Lambda_\textrm{QCD}=210\,\pm 14\,\textrm{MeV}$ \cite{PDG2016}, much larger than the  fundamental  quarks masses, clearly showing that the mass of the hadrons is due to the strong QCD dynamics.  Dark Energy corresponds to the lightest meson scalar particle $\phi$, dynamically formed due to the non-perturbative force of the DG. In the Minimal Supersymmetric Standard Model (MSSM)  the  gauge couplings are unified at the unification scale  $\Lambda_{\textrm{gut}}=(1.05\pm0.07)10^{16}\,\textrm{GeV}$ with  $g^2_{\textrm{gut}}= 4\pi/(25.83\pm 0.16)$ the coupling constant \cite{Amaldi}. As a natural extension we assume that our  DG is also unified with the SM gauge groups  and below this scale interact with the SM only via gravity. The DG gauge coupling evolves with energy and it becomes strong at  $\Lambda_c$, given by the one-loop renormalization equation \cite{Axel2003,Axel2005}:
\begin{equation} \label{eq:LambdacTh} 
\Lambda_c=\Lambda_{\textrm{gut}}\; e^{-8\pi^2/(b_og_{\textrm{gut}}^2)}=34 \,  ^{+16}_{-11}\,\,\textrm{eV},
 \end{equation}
where $b_o=3N_c-N_f=3$ is the one-loop beta function. Therefore, the condensation scale is not a free parameter of our model  but a  derived quantity.

At high energies ($T\gg 1\, \mathrm{TeV}$) all particles of the SM and DG are relativistic with energy densities $\rho_x = (\pi^2/30 )g_xT_x^4$ for $x$=SM,DG,  where   $g_\textrm{SM}^{gut}=228.75$  and  $g_\textrm{DG}^{gut}=97.5$ are the relativistic degrees of freedom for the MSSM and the DG, respectively \cite{Axel2003,Axel2005}. Since the SM and DG are unified at $\Lambda_{\textrm{gut}}$  the particles have the same temperature $T_{SM}^{gut}=T_{DG}^{gut}$ and the ratio of the energy densities is $\rho_\textrm{DG}^{gut}/\rho_\textrm{SM}^{gut} =g_\textrm{DG}^{gut}/g^{gut}_\textrm{SM}=0.43$.
Below $\Lambda_{\textrm{gut}}$ the SM and DG particles interact only via gravity and  are no longer maintained in thermal equilibrium.
We can relate the  temperatures using entropy conservation obtaining $T_{DG}/T_{\nu}=\le([g_\textrm{DG}^{gut}\, g_\textrm{SM}]/[g_\textrm{DG} \,g_ \textrm {SM}^{gut}] \ri)^{1/3}$ with $T_\nu$ the neutrino temperature.  The number of relativistic particles of the SM varies  with energy and at  neutrino decoupling  ($T\sim 1MeV$) we have $g_\textrm{SM}^{\nu dec}= 10.75$  while all DG particles remain  massless for $a\leq a_c$, giving $g_\textrm{DG}=g_\textrm{DG}^{gut}$.
At the phase transition $a_c$, which is below  neutrino decoupling,  we get the ratio:
\be\la{ocor}
\fr{\rho_\textrm{DG}^c}{\rho_\textrm{SM}^c} =\fr{g_\textrm{DG}^c}{g_\textrm{SM}^c}  \le(\fr{4}{11}\fr{g_\textrm{SM}^{\nu dec}}{g_\textrm{SM}^{gut}}\ri)^{4/3}= 0.1268,
\ee
with $\rho^c_\textrm{SM} = (\pi^2/30 )g^c_\textrm{SM}T_\gamma^4$ and $ g_\textrm{SM}^c=3.384$,  since at $a_c$ only photons and neutrinos remain relativistic, and  $T_\nu=(4/11)^{1/3} T_\gamma$.
Clearly, the DG amounts to a non-negligible fraction of the total relativistic energy content of the early universe. 
Extra relativistic particles beyond the SM are usually parameterised by the model independent quantity $N_{ex}$ given by $\rho_{ex}\equiv N_{ex}\,  (\pi^2/30) (7/4)\, T^4_\nu$. From eq.(\ref{ocor}) we obtain  $N_{ex}= (4/7) g_\textrm{DG}(g_\textrm{SM}^{\nu dec}/g_\textrm{SM}^{gut})^{4/3} =0.945$  for $a< a_c$   while  $N_{ex}=0$ for $a\geq a_c$ since  at $a_c$  all the  DG particles become massive due to the strong interaction of the DG \cite{Almaraz}. 

Once the condensation scale  $\Lambda_c$ is reached, the BDE meson fields $\phi$ are formed and we determine the scalar potential $V(\phi)$  using the analytical techniques studied in \cite{ADS}, giving an effective non-perturbative  IPL potential which is stable against radiative corrections \cite{Axel2003,Axel2005}:
\be
V=\Lc^{4+2/3}\phi^{-2/3},
\ee 
where the exponent of $\phi$ is given by $n=-2[1+2/(N_c-N_f)]=-2/3$. From dimensional analysis we set the physical quantities to be proportional to the symmetry breaking scale $\Lc$, giving the onset conditions of the BDE field $\phi(a_c)=\Lc$, $V(a_c)= \Lc^4$, $\rho_\textrm{DG}^c=2V(a_c)/(1-w_{\mathrm{BDE}}^c)=3\Lambda_c^4$, and $\dot{\phi}(a_c)=\sqrt{2\Lambda_c^4(1+w_{\mathrm{BDE}}^c)/(1-w_{\mathrm{BDE}}^c)}=2\Lc^2$, where $w_{\mathrm{BDE}}^c=1/3$ is the  EoS at $\ac$  and the dots stand for cosmic time derivatives. Setting $\rho_\textrm{SM}^{c}=\rho_\textrm{SM}^{o} a_c^{-4}$,  $g_\textrm{SM}^o=g_\textrm{SM}^c$, $T_{\gamma} ^o=2.7255\textrm{K}$ the present temperature of photons, we get  from eq.(\ref{ocor}): 
\be \label{eq:acLcTh}
\fr{a_c\Lc}{\textrm{eV}}=\le(\fr{\rho_\textrm{SM}^o}{3\textrm{eV}^4}\fr{g_\textrm{DG}^c}{g_\textrm{SM}^c} \ri)^{\fr{1}{4}} \le(\fr{4}{11}\fr{g_\textrm{SM}^{\nu dec}}{g_\textrm{SM}^{gut}}\ri)^{\fr{1}{3}}= 1.0939 \times 10^{-4},
\ee
which is a  meaningful prediction on the two essential parameters of BDE, subject to the constraint in eq.(\ref{eq:LambdacTh}).

\begin{table*}
  \caption{\small{Best fit (BF), mean and 68\% parameter CL. The ``base \LCDM'' parameters are given in rows $3-8$. $H_o$ is expressed in km$\cdot$s$^{-1}$Mpc$^{-1}$; $r_{\mathrm{BAO}}, f\sigma_8$, and $\gamma$ are evaluated at $z=0.57$.}}
\setlength{\extrarowheight}{1pt}
\begin{ruledtabular}
\begin{tabular}{l c c c c}
\multirow{2}{*}{Parameter} & \multicolumn{2}{c}{BDE} & \multicolumn{2}{c}{$\Lambda$CDM}\\
& best fit & 68\% limits & best fit & 68\% limits \\\hline
$\Lambda_c$ (eV) & 44.02 & 44.08 $\pm$ 0.27 & --- & ---\\
$10^{6}a_c$ & 2.48 & 2.48 $\pm$ 0.02 & --- & --- \\\hline
$\Omega_b h^2$ & 0.02252 & 0.02256$\pm$0.00021 & 0.02242 & 0.02238$\pm$0.00021\\
$\Omega_c h^2$ & 0.1173 & 0.1171 $\pm$ 0.0013 & 0.1181 & 0.1182 $\pm$ 0.0012\\
$100\theta_{\textrm{MC}}$ & 1.04106 & 1.04111$\pm$0.00042 & 1.04112 & 1.04112$\pm$0.00042\\
$\tau$ & 0.117 & 0.124 $\pm$ 0.027 & 0.118 & 0.110 $\pm$ 0.027\\
$10^9A_s$ & 2.37 & 2.40 $\pm$ 0.13 & 2.37 & 2.34 $\pm$ 0.12\\
$n_s$ & 0.9774 & 0.9780 $\pm$ 0.0049 & 0.9710 & 0.9701 $\pm$ 0.0048 \\\hline
$H_o$ & 67.68 & 67.80 $\pm$ 0.54 & 68.63 & 68.57 $\pm$ 0.58\\
$\Omega_{\textrm{DE}o}$ & 0.695 & 0.696 $\pm$ 0.007 & 0.702 & 0.701 $\pm$ 0.007\\
$w_{\textrm{DE}o}$ & -0.9296 & -0.9294 $\pm$ 0.0007 & $-1$ & $-1$\\
$\sigma_8(a_o)$ & 0.855 & 0.861 $\pm$ 0.022 & 0.871 & 0.864 $\pm$ 0.022\\
$r_{\rm{BAO}}$ & 0.07238 & 0.07247$\pm$0.00044 & 0.07230 & 0.07228$\pm$0.00043\\
$f\sigma_{8}$ & 0.4883 & 0.4909 $\pm$ 0.0124 & 0.5013 & 0.4978 $\pm$ 0.0123\\
$\gamma$ & 0.5500 & 0.5499 $\pm$ 0.0001 & 0.5492 & 0.5490 $\pm$ 0.0001\\
$z_{eq}$ & 3342 & 3339 $\pm$ 29 & 3359 & 3360 $\pm$ 29\\\hline
\multicolumn{5}{l}{$\chi^{2}_{\rm BDE} = 5.609\textrm{(BAO)} + 776.510\textrm{(CMB)} + 695.668\textrm{(SNeIa)} + 1.833\textrm{(prior)}$} \\
\multicolumn{5}{l}{$\chi^{2}_{\Lambda\mathrm{CDM}} = 7.115\textrm{(BAO)} + 776.883\textrm{(CMB)} + 695.075\textrm{(SNeIa)} + 1.681\textrm{(prior)}$}
\label{tab:paper_table} 
\end{tabular}
\end{ruledtabular}
\end{table*}

\begin{figure}[t]\centering \includegraphics[width=1\linewidth]{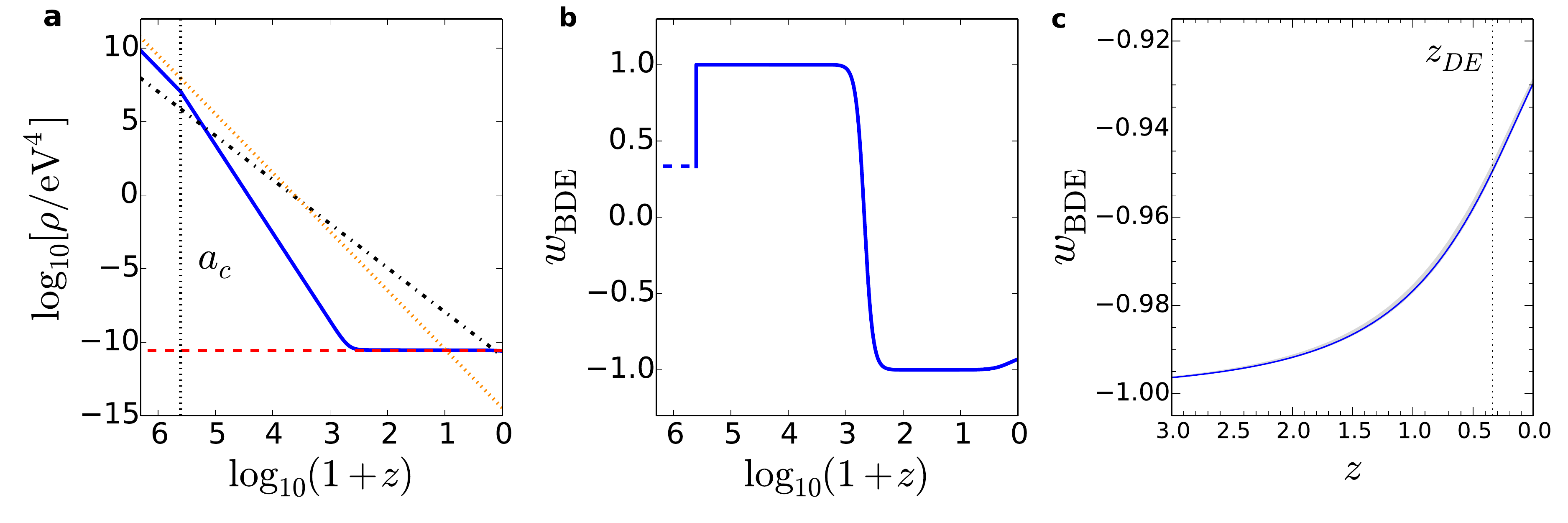}\caption{\small{(\textbf{a}) $\rho_{\textrm{BDE}}$ (blue, solid), $\rho_{m}$ (black, dash-dotted), $\rho_{r}$ (orange, dotted), and $\rho_{\Lambda}$ (red, dashed). (\textbf{b}) BDE EoS for $a<\ac$ (dashed) and $a\geq \ac$ (solid). (\textbf{c}) BDE EoS at late times. $z_{DE}$ marks the matter-dark enery equality epoch.}}
\label{fig:paper_figure1}
\end{figure}

The evolution of BDE field in a homogeneous flat universe described by the Friedmann-Lema\^itre-Robertson-Walker metric is completely determined by the Klein-Gordon $\ddot{\phi}+3H\dot{\phi}+dV/d\phi=0$ and Friedmann $H\equiv \dot a/a= \sqrt{8\pi G \rho_{tot}/3}$ equations.
 The total energy density is $\rho_{tot}(a)=\rho_{mo}a^{-3}+\rho_{ro}a^{-4}+\rho_{\mathrm{BDE}}$  with
 $\rho_{\mathrm{BDE}}=\dot\phi^2/2+V(\phi)$ for BDE and $\rho_{mo}$,   $\rho_{ro}$ are the present day   matter and radiation densities, while the redshift $z$ is given by $a=(1+z)^{-1}$ and $a_o=1$ at present time. 
In the  standard $\Lambda$CDM model  $\rho_{mo}$ and the size of the cosmological constant $\rho_\Lambda=\Lambda$  are  free parameters to be determined by observations. However, in BDE for given $\rho_{mo}$ the value of $\rho_{\mathrm{BDE}}$ is a predicted quantity given by the solution of the Klein-Gordon and Friedmann equations whose initial conditions are fully specified as we have just seen. Therefore, BDE not only has no free parameters but also posses one less  than \LCDM.

We study  the cosmological implications of our  BDE model  and compare them with $\Lambda$CDM to highlight the differences. For that purpose we perform a Markov Chain Monte Carlo MCMC analysis using the \texttt{CosmoMC} \cite{CosmoMC} and \texttt{CAMB} \cite{CAMB} codes properly adapted to describe the full background and linear perturbation dynamics. 
We consider measurements of the CMB temperature anisotropies \cite{Planck2015}, BAO \cite{GilMarin,Ross,Beutler}, and SNeIa \cite{Betoule} data. 
We vary $\Lambda_c$  and determine $a_c$ from the constriction given by eq.(\ref{eq:acLcTh}). 
Table \ref{tab:paper_table} quotes the best fits (BF) with their corresponding g.o.f. ($\chi^2$) and the mean and 68\% CL of some selected parameters. For our BDE model we obtain $\Lambda_c(\mathrm{eV})= 44.08 \pm 0.27$, $a_c=(2.48 \pm 0.02)\times 10^{-6}$, $\Omega_{\mathrm{BDE} o}=0.696 \pm 0.007$, and  $w_{\mathrm{BDE} o}=-0.929 \pm 0.001$.  
Notice that  BDE has an excellent agreement with the cosmological measurements and a better fit than \LCDM\, even though it has one less free parameter. Specifically, BDE has a significant improvement  of  $\chi^2_{BAO}$ by 21\%, showing that a dynamical DE is preferred.  All  base \LCDM\,  parameters \cite{Planck2015} are consistent within $1\sigma$ with  BDE. However,  we find relevant tensions at more than $2\sigma$ between BDE and \LCDM\,  for BAO measurements and structure growth.
We also test our theoretical constriction of eq.(\ref{eq:acLcTh}) by allowing $a_c$ and $\Lc$ to vary freely  and independently and we  find remarkable  that the relative difference  between eq.(\ref{eq:acLcTh}) with the  best-fit value $a_c\Lambda_c/\eV=1.0916\times 10^{-4}$  is only 0.2\%. 
The evolution of the different components for the BF is shown in (\ref{fig:paper_figure1}\textbf{a}). Notice that at early times $\rho_{\Lambda} \ll \rho_r$ exposing the naturalness and coincidence problems of the \LCDM\, model. Since for $a<a_c$ our model   contains relativistic particles, its energy density evolves as $\rho_\textrm{BDE}\propto a^{-4}$  with non negligible energy  density $ \Omega_{DG}/\Omega_{SM}=0.43 ( 0.13) $ for $a=a_{gut}(a_{c})$, respectively.
However, at $a_c$ the phase transition takes place and BDE dilutes rapidly as  $\rho_\textrm{BDE}\propto a^{-6}$, taking its  minimum value  
 $\Omega_\textrm{BDE}(a\simeq 10^{-3})\simeq 10^{-8}$  and  it becomes dominant at late times with  $\rho_\textrm{BDE}\approx const $
and  $\Omega_\textrm{BDE}\simeq 0.69 $ at present time.  

\begin{figure}[t]\centering \includegraphics[width=1\linewidth]{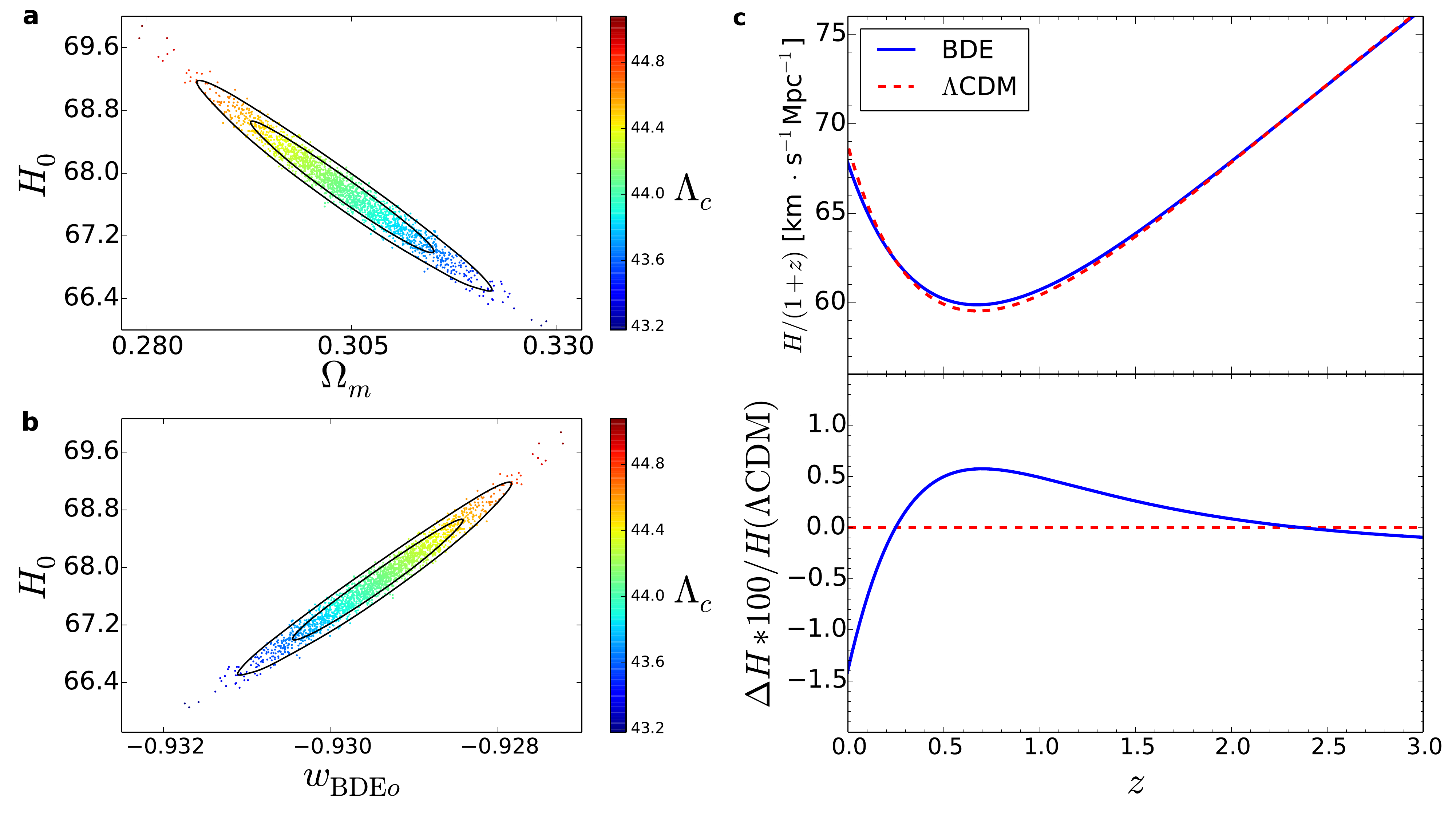}\caption{\small{(\textbf{a, b}) Samples in the $H_o$-$\Omega_m$ and $H_o$-$w_{\mathrm{BDE} o}$ planes coloured by the condensation scale $\Lambda_c$ (in eV). The contours mark the 68\% and 95\% CL. (\textbf{c}) Hubble expansion rate. The lower panel displays the relative difference w.r.t. $\Lambda$CDM.}}
\label{fig:paper_figure2}
\end{figure}

In Figs.(\ref{fig:paper_figure1}\textbf{b}) and (\ref{fig:paper_figure1}\textbf{c})  we show the EoS of BDE and we notice that after $\ac$ $w_{\mathrm{BDE}}$ leaps to $1$ and remains at this value for a long  period of time, then drops to $-1$ shortly after decoupling $z_*=1089.98$  to finally grow to $w_{\textrm{BDE} o} = -0.93$ at present time.  
For the BF we obtain the bounds  $-1< w_{\textrm{BDE}} \leq -0.99,-0.95$ for $z\geq 1.8, 0.35$ while $\Omega_{\textrm{BDE}} \leq 1\%, 0.1\%$ for $z\geq 5.3, 12.7$.
Figs.(\ref{fig:paper_figure2}\textbf{a}) and (\ref{fig:paper_figure2}\textbf{b}) shows the impact of $\Lambda_c$ on the present density matter ($\Omega_m$), the current expansion rate ($H_o$), and the BDE EoS. We see that larger values of $\Lambda_c$ lead to smaller values of $\Omega_m$ and larger $H_o$ and $w_{\mathrm{BDE} o}$, this latter being tightly constrained.

\section{Observational Constraints}

\subsection{ Distances and BDE Late-Time Dynamics}

The behaviour of the equation of state at recent times leads to a distinctive late-time dynamics (LTD) which has a broad impact on the cosmological observables since it modifies the amount of DE at late times. This is specially manifest in cosmological distances probed by SNIa, BAO, and CMB measurements as well as in the evolution of matter perturbations and CMB anisotropies. Distances are affected by  the size and evolution of $H(z)$ which in turn depends on the amount of DE, and since $w_\textrm{BDE}> -1$ at late times, $\rho_\textrm{BDE}$ increases as a function of $z$ while $ \rho_{\Lambda}$ remains constant, so we expect to see differences in BDE and \LCDM. 

\begin{figure}[t]\centering \includegraphics[width=0.85\linewidth]{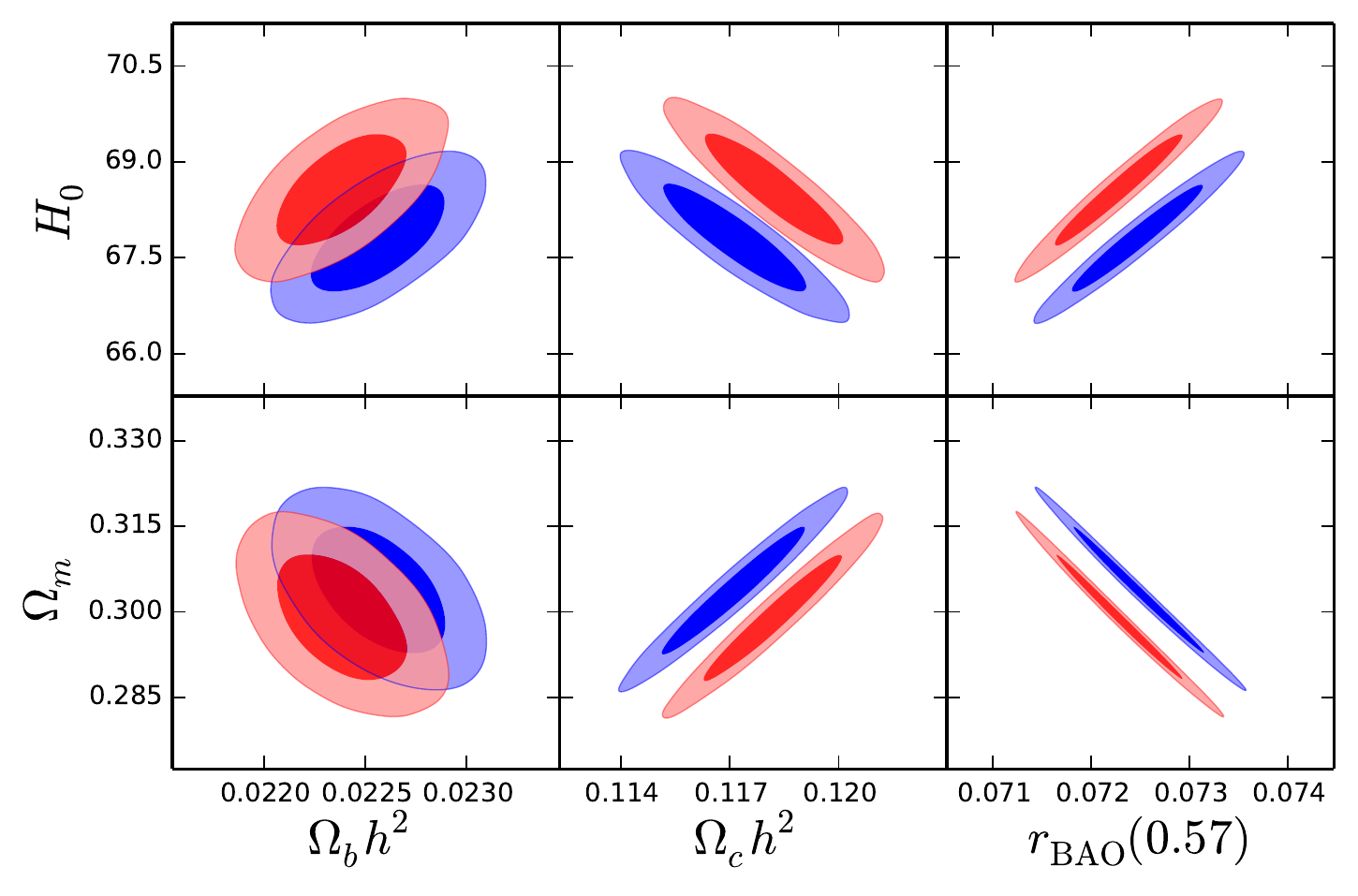}\caption{\small{68\% and 95\% CL contours of $H_0$ and the present matter density $\Omega_m$ vs the baryon and CDM physical densities, and the BAO ratio at $z=0.57$ for the BDE (blue) and $\Lambda$CDM (red) models.}}\label{fig:paper_figure3}
\end{figure}

Fig.(\ref{fig:paper_figure2}\textbf{c}) shows the deviation of the expansion rate of BDE with respect to \LCDM\, for the BF.  We see that  $H(z)$ is larger in BDE than in \LCDM\, in the range $0.24<z<2.3$ sensitive to BAO and SNIa measurements with discrepancy of up to $0.58\%$ at $z=0.7$. On the other hand, $H_o$  is smaller in  BDE  than  in \LCDM\,. This is because the accurate determination of the angular size of the sound horizon at recombination obtained from CMB measurements forces BDE and \LCDM\, to  have the same angular distance $D_A(z_*)$ (a difference by less than $0.05\%$), and  since the amount of matter is roughly the same (difference less than $0.5\%$) the  amount of $\rho_\textrm{BDE}$ at present time  must be smaller than $ \rho_{\Lambda}$ (it is  $3.7\%$ smaller), giving a lower value of $H_o^2$ in BDE than in \LCDM\, by 2.75\%. Even though the base \LCDM\,  parameters  are consistent within $1\sigma$,  we see  in  Fig.(\ref{fig:paper_figure3})  a tension at more  than $2\sigma$  in plots of $H_o$($\Omega_m$) vs $\Omega_c h^2$ and  $H_o$($\Omega_m$) vs $r_{\textrm{BAO}}\equiv r_{\textrm{drag}}/D_V$ at $z=0.57$, where  $D_V(z)=[(1+z)^2D_A(z)^2z/H(z)]^{1/3}$ with $D_A(z)=(1+z)^{-1}\int_o^{z} dz/H(z)$ and $r_{\textrm{drag}}$ the comoving sound horizon at the drag epoch \cite{Planck2015}. These combinations of parameters allow us to probe the dynamics of the DE and is precisely in the BAO ratio $r_{\textrm{BAO}}$  where we obtain a 20\% reduction in $\chi^2_{BAO}(z= 0.57)$ for BDE, favouring our dynamical DE model.

The background evolution of the BDE scalar field  $\phi$ can be well approximated by the EoS   $w_\textrm{fit}= (-0.929 - 3.752\, z - 5.926\, z^2 - 4.022\, z^3 - 0.999\, z^4) / (1 + z)^4$  with a relative error with  $w_\textrm{BDE}$ below $0.1\%$  for  $z< 140$ (see appendix \ref{App}) and therefore the cosmological distances  remain unchanged.

\subsection{Matter Power Spectrum}\label{SecMPW}
\begin{figure*}\centering \includegraphics[width=1.4\columnwidth]{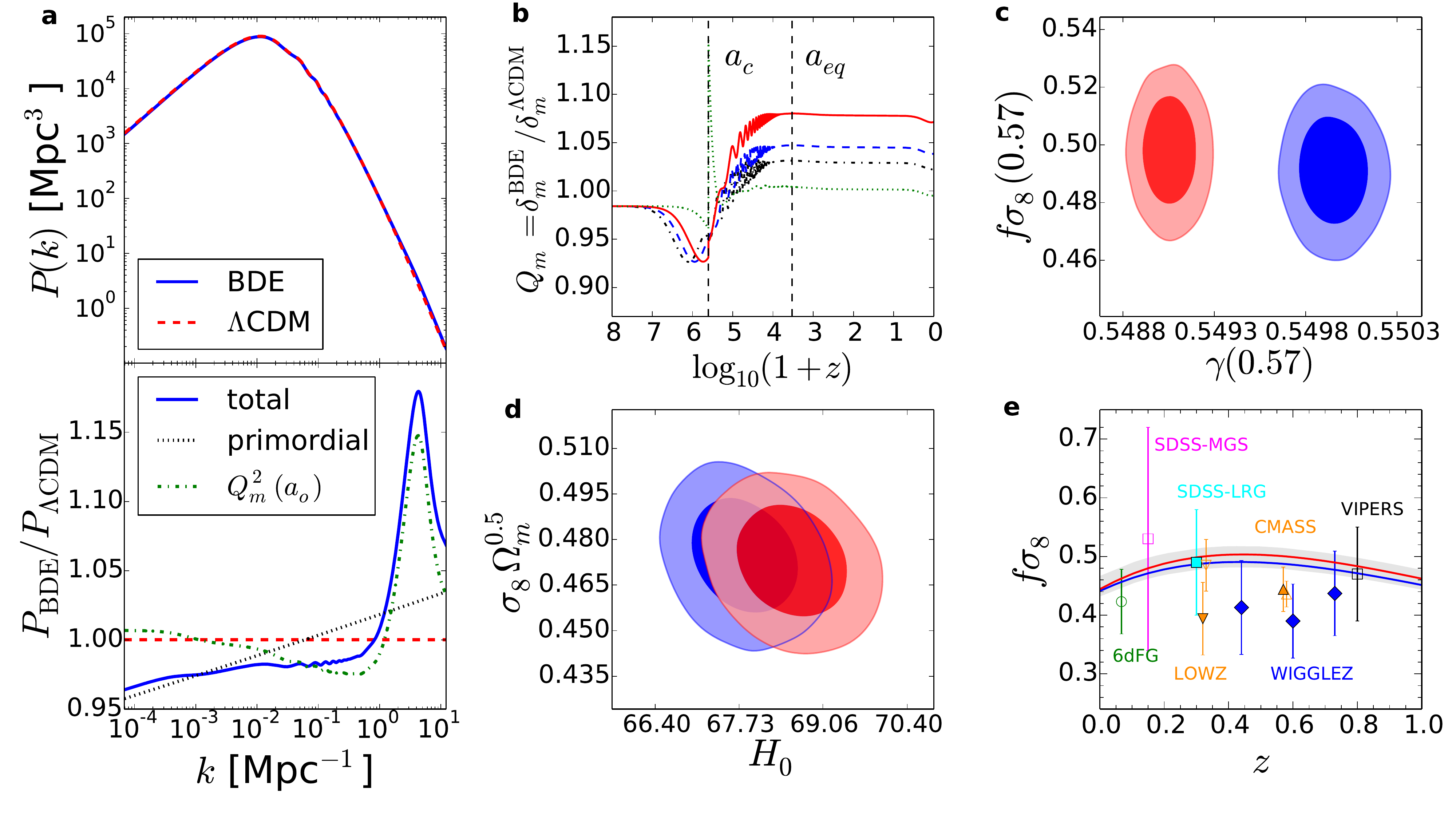}\caption{\small{(\textbf{a}) Matter power spectrum for the best fit. The lower panel shows the ratio w.r.t. $\Lambda$CDM of the total spectrum (blue solid), the primordial spectrum (black dotted), and $|\delta_m|^2 $ at present time (green, dash-dotted). (\textbf{b}) Ratio of $\delta_m$ (in the newtonian gauge) for $k=1$ (green, dotted), $4.3$ (red, solid), $7$ (blue, dashed), and $10\textrm{Mpc}^{-1}$ (black, dash-dotted). (\textbf{c}, \textbf{d}) 68\% and 95\% CL contours in the $\gamma-f\sigma_8$ at $z=0.57$ and $H_0-\sigma_8\Omega_m^{0.5}$ planes for BDE (blue) and $\Lambda$CDM (red). (\textbf{e}) Constraints on $f\sigma_8$ for BDE (blue) and $\Lambda$CDM (red). The grey band marks the 95\% CL for BDE allowed by the datasets analysed in this work . The dots are the measurements of some galaxy surveys (see references).}}
\label{fig:paper_figure4}\end{figure*}

The overall dynamics of the dark energy in the BDE model leaves important imprints on the evolution of matter perturbations $\delta_m\equiv \delta\rho_m/\rho_m$. Small modes  ($k>k_c\equiv\ac H_c=0.925 \textrm{Mpc}^{-1}$)  entering the horizon before $\ac$ have  distinctive features in BDE compared to standard \LCDM\, as shown in Fig.(\ref{fig:paper_figure4}\textbf{b}). Initially, the extra free streaming particles $N_{ex}=0.945$ of the DG suppress the matter perturbations with respect to $\Lambda$CDM by nearly 1.6\%, i.e., $\delta_{mi}^{BDE}/\delta_{mi}^{\Lambda CDM}\approx 0.984$. This suppression is model independent and cannot be compensated by varying other cosmological parameters \cite{HuSugiyama}. 
The difference of the scale factor at  horizon crossing is given by $a_{h}^{BDE}/a_{h}^{\Lambda CDM} =\sqrt{1+\rho_{DG}^c/\rho_{SM}^c}=1.062$ (c.f. eq.(\ref{ocor})) allowing more time for $\delta_m$ to grow in \LCDM\, and suppressing  the BDE modes further. 
However,  the change in the expansion rate after the rapid dilution of BDE makes the matter perturbations in BDE grow at a higher rate which not only compensates but  reverses the initial suppression of the first two effects. The enhancement is mode dependent reaching a maximum of 7\% for $k\approx 4.3\textrm{Mpc}^{-1}$, which agrees with the semianalytical estimation $\delta_m^{BDE}/\delta_m^{\Lambda CDM}-1=  (\delta_{mi}^{BDE}/\delta_{mi}^{\Lambda CDM})(H_+^B/H_-^B)-1\simeq 5\%$ valid for modes $k> k_c$ in the range $a_{eq}> a\gg \ac$, where $H_+^B/H_-^B  =\sqrt{1+\rho_{DG}^c/\rho_{\mathrm{SM}}^c}$ and $a_{eq}$ is the matter-radiation equality epoch \cite{Almaraz}.
During the matter domination era $\delta_m$ grows $\propto a$ for all modes both in BDE and  $\Lambda$CDM. However, at late times the LTD and the BDE field inhomogeneities suppress the growth rate of matter perturbations. The suppression factor is nearly the same for all the modes, giving a drop of $\Delta\delta_{m}=-0.61\%$ for the BF, with equal contributions from the LTD of the background and the DE perturbations.

The final shape of the matter power spectrum $P= 2\pi^2 P_s\, |\delta_{m}(a_o)|^2/k^3$  is a combination of the present value of $\delta_m(a_o)$ determined by the processes described above and the best fit values of $n_s$ and $A_s$ which define the primordial spectrum $P_s\equiv A_s(k/k_o)^{n_s-1}$. In Fig.(\ref{fig:paper_figure4}\textbf{b}) we show the  the differences in the spectra for the BF. The different tilt $n_s^{\mathrm{BDE}}>n_s^{\Lambda\mathrm{CDM}}$ suppress the spectrum for large modes ($k<k_o=0.05 \textrm{Mpc}^{-1}$) and enhances it for $k>k_o$ in the BDE model. We obtain a suppression of only 1 to 3\% for modes $k< k_c$ while the net effect for modes $k> k_c$ is an increase of up to 18\% for BDE peaking at  $k\approx 4.3 \textrm{Mpc}^{-1}$, where the effect of the rapid dilution is maximum.
This scale corresponds to a structure of radius $r=\pi/k=0.7\textrm{Mpc}$ with a mean mass $M=(4\pi/3) r^3 \rho_{mo}=6.3\times 10^{10} M_{\odot}$  at present time. In this regard, the  enhancement in the power spectrum also increases the number density of galaxies of different sizes $dn/dlog$. We have seen that the rapid dilution of DE strongly affects the evolution of modes in the range  $0.6\textrm{Mpc}^{-1}< k < 9.4\textrm{Mpc}^{-1}$ corresponding to radiuses between $5\textrm{Mpc}>r>0.3\textrm{Mpc}$.  Using the Press-Schechter mass function \cite{PressSchechter}, we find an increase of 4\% in the number density $dn/dlog$ for masses between  $M_i=(9\times 10^{9}-1\times10^{14})M_{\odot}$ compared to \LCDM\,.
However, the  final results  depend  on the properties and amount of DM  and since modes $k>k_c$ are  no longer in the linear regime a  non-linear approach  must be used.

The imprints on the structure formation can also be observed in the growth index $\gamma$ and $f\sigma_8$, where $f\equiv d ln (\delta_m )/d ln(a)=\Omega_m^\gamma(a)$. While fig. (\ref{fig:paper_figure4}\textbf{c}) shows a clear tension with $\Lambda$CDM in the $\gamma-f\sigma_8$ plane at $z=0.57$, other parameter combinations such as $H_0-\Omega_m^{0.5}\sigma_8$ in fig.  (\ref{fig:paper_figure4}\textbf{d}) are consistent at the $1\sigma$ level. For the BF, the differences in $\gamma$ are lower than 0.3\%, while the deviations in $f\sigma_8$ in fig. (\ref{fig:paper_figure4}\textbf{e}) are up to 2.6\% in the region $0.4\leq z \leq 0.8$ with BDE suppressing $f\sigma_8$ by  2.3\% (0.45\%) at $ z=0\,(0.57)$ \cite{Almaraz}. Future studies on redshift-space distortions will provide key evidence to settle this issue \cite{Beutlerfs8,Howlettfs8,Okafs8,GilMarinfs8,Blakefs8,DelaTorrefs8}.

\subsection{ Extra Relativistic Particles}
The presence of extra relativistic degrees of freedom can be constrained by current cosmological observations, so in order to be a viable model of dark energy our BDE model must be in agreement with these constraints. The amount of radiation besides photons is usually parametrized by $N_{\textrm{eff}}\equiv N_\nu+N_{ex}$, where $N_\nu=3.046$ for 3 massless neutrino species. Standard analyses consider a constant $N_{\textrm{eff}}$ over the whole history of the universe (e.g., \cite{Planck2015}). 
These extra relativistic particles increase the expansion rate at early times modifying the amount of primordial elements formed at BBN. They affect the damping tail of the CMB spectrum \cite{HuWhite,Hou} and shift the matter-radiation epoch to a later time, leaving an additional imprint on the CMB which can be probed by the early Integrated Sachs-Wolfe effect \cite{Almaraz,eISW}. Extra relativistic particles also introduce additional anisotropic stress and modify the evolution of radiation and matter anisotropies.
In our BDE  model $N_{ex}$ changes from $N_{ex}=0.945 $ for $a<\ac$ to $N_{ex}=0$ for $a\geq \ac$, leaving then the matter-radiation equality and recombination epochs unchanged. Therefore BDE describes a cosmological scenario different than the usual constant $N_{\textrm{eff}}$.
However, the extra amount of radiation in BDE during BBN increases the primordial helium $\textrm{Y}_{P}$ and deuterium ($D/H$) abundances too. 
For the BDE model, we obtain \small{$\textrm{Y}_{P}=0.2587\pm 0.0001\, (0.0003)$}  \normalsize{and} \small{$\textrm{D/H}=(2.88\pm 0.046 \, (0.06))\times 10^{-5}$}\normalsize{ at $68\%$ CL}. Although the precise BBN  abundances are still under investigation and have significant uncertainties due to the cosmological measurements and the neutron life time \cite{BBNOlive}, these results are consistent with the abundances obtained by astrophysical probes \cite{Aver2013,Izotov2014,Iocco2009,Riemer2015} well within the $2\sigma$ level.

\section{Conclusions}
We have seen that our BDE model is a natural extension of the SM of particles and  without introducing any free parameters we are able to understand the current  acceleration of the universe due to the dynamics of a light dark meson field. BDE  also describes extra relativistic particles at high energies and a  rapid dilution of its energy density at $\ac$.  All these a  priori unconnected phenomena leave distinctive measurable imprints in the universe.  Our BDE model is not only predictive but  it allows to understand the nature of DE.  \\

\appendix
\section{Equation of state fit}\label{App}

\begin{figure}[t]\centering \includegraphics[width=1\linewidth]{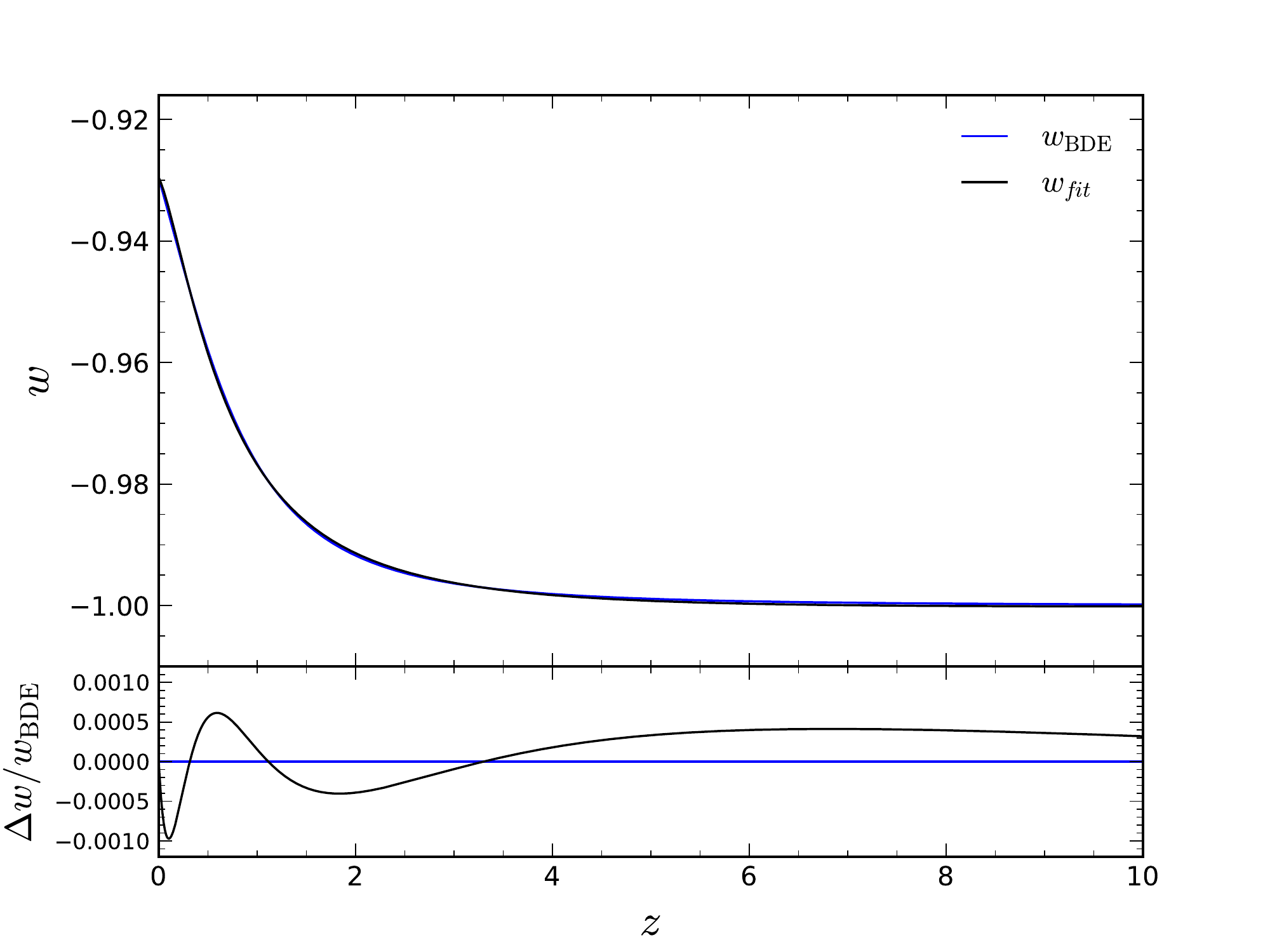}\caption{We show $w_\textrm{BDE}$ and $w_\textrm{fit}$  and the relative difference $\Delta w= (w_\textrm{fit}-w_\textrm{BDE})/w_\textrm{BDE}$  as function of the redshift $z$. }\label{fig:paper_figura5}
\end{figure}

Instead of solving the dynamical  equation  of  the BDE  background    given by scalar field  $\phi$,  we can estimate its  evolution and cosmological effects  by using an effective EoS given by the ansatz   $w_\textrm{fit}= (-0.929 - 3.752\, z - 5.926\, z^2 - 4.022\, z^3 - 0.999\, z^4) / (1 + z)^4$. We show  in Fig.(\ref{fig:paper_figura5}) the evolution of $w_\textrm{fit}$  and compared it to $w_\textrm{BDE}$, obtaining an  excellent fit with a relative error  below  $0.1\%$  valid for  $z< 140$,  before $w_\textrm{BDE}(z)$ starts to  grow to $w_\textrm{BDE}=1$  (see Fig.(\ref{fig:paper_figure1}b)).  Our EoS ansatz accounts then  for the background evolution and  gives  equivalent cosmological distances and suppression factor of the matter perturbations as our BDE model.  The BDE perturbations are not accounted for in  our anstaz $w_\textrm{fit}$  of the DE background, however its  contributions  to the linear growth of matter perturbations are smaller than $1\%$ (see section \ref{SecMPW}) and they do not affect the cosmological distances.

\bibliography{BoundDarkEnergy_RevTex_vax} 

\end{document}